\documentclass[journal]{IEEEtran}

\usepackage{graphicx}
\usepackage{amsmath}
\usepackage{amssymb}
\usepackage{amsthm}
\usepackage{epsfig}
\usepackage[utf8]{inputenc}
\usepackage[T1]{fontenc}
\usepackage{cite}
\usepackage{fixltx2e}
\usepackage{balance}
\usepackage{psfrag}
\usepackage{balance}
\usepackage{color}

\newcommand{\vv}[1]{\mathbf{#1}}
\newcommand{\mm}[1]{\mathbf{#1}}

\newcommand{\expect}[1]{{\mathbb{E}}\!\left\{ #1 \right\}}

\newcommand{\herm}{{\sf{H}}}
\newcommand{\transp}{{\sf{T}}}
\renewcommand{\phi}{\varphi}

\renewcommand{\epsilon}{\varepsilon}

\ifCLASSINFOpdf
\else
\fi

\begin{document}
%
\title{Low-Complexity Widely-Linear Precoding for Downlink Large-Scale MU-MISO Systems}
%
%
%
\author{Shahram~Zarei,~\IEEEmembership{Student Member,~IEEE,}
        Wolfgang~Gerstacker,~\IEEEmembership{Senior Member,~IEEE,}
        and~Robert~Schober,~\IEEEmembership{Fellow,~IEEE}
\vspace*{-7mm}        
}

%
%

\markboth{}%
{Shell \MakeLowercase{\textit{et al.}}: Bare Demo of IEEEtran.cls for Journals}
%



\maketitle

\begin{abstract}
In this letter, we present a widely-linear minimum mean square error (WL-MMSE) precoding scheme employing
real-valued transmit symbols for downlink large-scale multi-user multiple-input single-output (MU-MISO) systems. In contrast to
the existing WL-MMSE transceivers for single-user multiple-input multiple-output (SU-MIMO) systems, which use both WL
precoders and WL detectors, the proposed scheme uses WL precoding only and simple conventional detection at the user
terminals (UTs). Moreover, to avoid the computational complexity associated with inversion of large matrices, we modify the
WL-MMSE precoder using polynomial expansion (PE). Our simulation results show that in overloaded systems, where the
number of UTs is larger than the number of base station antennas, the proposed PE WL-MMSE precoder with only a few
terms in the matrix polynomial achieves a substantially higher sum rate than systems employing conventional MMSE precoding.
Hence, more UTs sharing the same time/frequency resources can be served in a cell. We validate our simulation results with
an analytical expression for the asymptotic sum rate which is obtained by using results from random matrix theory.
\end{abstract} 
%
\IEEEpeerreviewmaketitle

%
\vspace*{-4mm}
\section{Introduction}
\IEEEPARstart{M}{ultiple-}input multiple-output (MIMO) technology enables a substantial increase in spectral efficiency and transmission reliability in wireless communication systems. An emerging research field in MIMO communications are so-called large-scale MIMO systems, where base stations are equipped with a large number of antennas, e.g., hundred or more. Large-scale MIMO systems enable very high spectral and power efficiencies \cite{Rusek2013}.

In this letter, we consider the downlink (DL) of a large-scale multi-user multiple-input single-output (MU-MISO) system, which embodies a Gaussian broadcast channel (GBC). It is known that for the GBC, nonlinear dirty paper coding (DPC) is capacity achieving \cite{Caire2003}. However, due to the high computational complexity of DPC, linear precoding schemes such as minimum mean square error (MMSE) precoding are attractive alternatives. Moreover, in the asymptotic scenario, where the number of base station antennas, $N,$ and the number of user terminals (UTs), $K,$ are very large but $K$ is significantly smaller than $N$, the linear MMSE precoder with complex Gaussian transmit symbols achieves near optimum performance in terms of the sum rate \cite{Rusek2013}. 
On the other hand, in DL MU-MISO systems, if the number of UTs, $K$, is very large, and $K$ is much larger than the number of base station antennas, $N$, so-called semi-orthogonal user selection zero-forcing (SUS-ZF) precoding achieves the same asymptotic sum rate as DPC \cite{Yoo2006}. 

For code division multiple access (CDMA) systems and single-user MIMO (SU-MIMO) systems with improper transmit symbols, i.e., transmit symbols with non-zero pseudo-covariance, it has been shown that so-called widely-linear MMSE (WL-MMSE) detectors outperform conventional MMSE detectors \cite{Lampe2001}. In a WL-MMSE detector, both the received signal and its complex conjugate are filtered separately and independently, and the filter outputs are combined \cite{Lampe2001}, \cite{Sterle2007}, \cite{Darsena2013}. 
Recently, the authors of \cite{Darsena2013} introduced a joint optimization approach for designing WL precoders and detectors for SU-MIMO systems. However, in MU-MISO systems, due to their decentralized structure, the application of WL detectors such as those proposed in \cite{Darsena2013} is not possible. In this letter, we propose a WL precoding scheme for MU-MISO systems, which uses real-valued data symbols and does not require any signal processing at the UTs. The goal of the optimization is the minimization of the sum mean square error (sum MSE) between the real part of the received symbols and the real-valued data symbols under a sum transmit power constraint. 
However, the obtained solution still entails a high computational complexity due to the required inversion of a large matrix. To overcome this problem, we exploit the large system properties of large-scale MU-MISO systems and extend the results of \cite{Zarei_PIMRC2013} to approximate the matrix inversion in the WL-MMSE precoder by a matrix polynomial. Finally, using results from random matrix theory, we obtain analytical expressions for the asymptotic signal-to-interference-plus-noise ratio (SINR) and the asymptotic sum rate. 

The contributions of this letter are summarized as follows. First, we develop a WL-MMSE precoder for real-valued transmit symbols and show that it yields a substantially higher sum rate than the commonly used MMSE precoder, when the number of UTs is larger than the number of base station antennas. This is different from the work in \cite{Shi2007}, where a framework for calculation of strictly linear MMSE downlink transceiver filters from uplink filters was introduced. Second, in contrast to the existing WL-MMSE transceivers, where signal processing is performed both at the transmitter and the receiver, in our proposed scheme, signal processing at the receiver is not required. This makes the proposed scheme attractive for decentralized applications, i.e., MU-MISO systems. Third, using results from random matrix theory, we also propose a polynomial expansion (PE) WL-MMSE precoder, which is based on a matrix polynomial instead of matrix inversion and further reduces the computational complexity. Fourth, our numerical results show that the proposed PE WL-MMSE precoder achieves a sum rate which is very close to the sum rate of the SUS-ZF precoder proposed in \cite{Yoo2006} but entails a lower computational complexity. We consider SUS-ZF as a performance benchmark because of its excellent performance, when the number of UTs is larger than the number of base station antennas.

\emph{Notation:} Boldface lower and upper case letters represent column vectors and matrices, respectively. $\mathrm{diag} \left(Q_1, \dots, Q_K\right)$ is a diagonal matrix with scalars $Q_1,\dots,Q_K$ on its main diagonal. $\mm{I}_K$ denotes the $K \times K$ identity matrix and ${\left[\mm{A}\right]}_{m,:}$, ${\left[\mm{A}\right]}_{:,n}$, and ${\left[\mm{A}\right]}_{m,n}$ stand for the $m$th row, the $n$th column, and the element in the $m$th row and the $n$th column of matrix $\mm{A}$, respectively. $(\cdot)^*$ denotes the complex conjugate and $\mathrm{tr}(\cdot)$, $(\cdot)^{\transp}$, and $(\cdot)^{\herm}$ are the trace, transpose, and Hermitian transpose of a matrix, respectively. $\Re\lbrace \cdot \rbrace$ stands for the real part of a complex variable and $\Vert \vv{a} \Vert$ represents the Euclidean norm of vector $\vv{a}$. $\expect{\cdot}$ refers to the expectation operator and $\mathcal{C} \mathcal{N} \left(\vv{m}, \mm{\Phi} \right)$ denotes a circular symmetric complex Gaussian distribution with mean vector $\vv{m}$ and covariance matrix $\mm{\Phi}$.
\vspace*{-4mm}
\section{System Model}\label{Sec_SystemModel}
\begin{figure}
\begin{center}
\psfrag{d}[cc][cc][1]{$\vv{d}$}
\psfrag{x}[cc][cc][1]{}
\psfrag{V}[cc][cc][1]{$\tilde{\mm{V}}$}
\psfrag{sQ}[cc][cc][1]{$\ {\mm{P}}^{1/2}$}
\psfrag{s}[cc][cc][1]{}
\psfrag{G}[cc][cc][1]{$\tilde{\mm{H}}$}
\psfrag{n1}[cc][cc][1]{$n_{\mathrm{R}_1}$}
\psfrag{nK}[cc][cc][1]{$n_{\mathrm{R}_K}$}
\psfrag{x1}[cc][cc][1]{}
\psfrag{xK}[cc][cc][1]{}
\psfrag{r1}[cc][cc][1]{}
\psfrag{rK}[cc][cc][1]{}
\psfrag{r}[cc][cc][1]{$\vv{r}$}
\psfrag{dl1}[cc][cc][1]{$\delta_1$}
\psfrag{dlK}[cc][cc][1]{$\delta_K$}
\psfrag{q1}[cc][cc][.7]{$P_1^{-1/2}$}
\psfrag{qK}[cc][cc][.7]{$P_K^{-1/2}$}
\psfrag{d1}[cc][cc][1]{$\hat{d}_1$}
\psfrag{dK}[cc][cc][1]{$\hat{d}_K$}
\includegraphics[width=\linewidth]{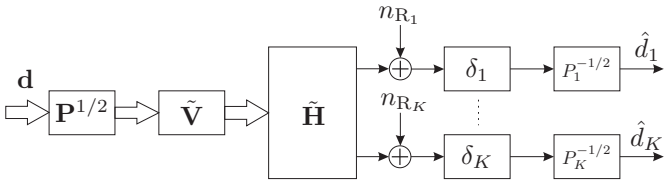}
\caption{Downlink augmented real-valued system model.}
\label{fig:DL_BD}
\end{center}
\vspace*{-6mm}
\end{figure}

We consider the downlink of a single-cell large-scale MU-MISO system, where a base station with $N$ antennas transmits signals to $K$ single-antenna UTs which are randomly and uniformly distributed within the cell. Each UT occupies the same time and frequency resources. $N$ and $K$ are assumed to be large with their ratio $\beta=K/N$ being constant. We consider a flat fading channel, and we further assume that the channel state information (CSI) is perfectly known at the transmitter. The real-valued, independent and identically distributed (i.i.d.) zero-mean unit-variance Gaussian data symbols of the $K$ UTs are stacked into  vector $\vv{d} =\left[ d_1 \ldots d_K \right]^\transp \in \mathbb{R}^K$ with  $\expect{\vv{d} {\vv{d}}^\transp}={\vv{I}}_K$. \footnote[1]{Throughout this letter, we assume real-valued Gaussian data symbols for WL-MMSE precoding, whereas for conventional ZF and MMSE precoding, which are considered as benchmark schemes, complex-valued Gaussian data symbols are assumed as usual.}
The vector of the stacked detected symbols of all UTs is given by 
\vspace*{-2mm}
\begin{equation}
\hat{\vv{d}} = \Re \left\lbrace \mm{P}^{-1/2} \mm{\Delta} \mm{H} \mm{V} \mm{P}^{1/2} \vv{d} + \mm{P}^{-1/2} \mm{\Delta} \vv{n} \right\rbrace,
\vspace*{-2mm}
\end{equation}
where channel matrix $\mm{H} $ models i.i.d. Rayleigh fading with ${[\mm{H}]}_{m,n} \sim \mathcal{C} \mathcal{N} \left(0,1\right)$. $\mm{V} \in \mathbb{C}^{N \times K}$ is the normalized precoding matrix with unit norm columns. $\mm{\Delta}=\mathrm{diag} \left( \delta_1,\dots,\delta_K \right)$ contains real-valued scaling factors for all UTs and $\mm{P} = \mathrm{diag} \left( P_1,\dots,P_K \right)$ is the power allocation matrix with $P_i$ being the $i$th UT's transmit power. $\vv{n} =\left[ n_1 \ldots n_K \right] ^\transp \sim \mathcal{C} \mathcal{N} \left(\vv{0}, \sigma_n^2 {\vv{I}}_K\right) $ is an additive white Gaussian noise (AWGN) vector whose entries have variance $\sigma_n^2$. Using augmented real-valued vectors and matrices, $\hat{\vv{d}}$ can be equivalently expressed as
\vspace*{-1.2mm}
\begin{equation}
\hat{\vv{d}} = \mm{P}^{-1/2} \mm{\Delta} \tilde{\mm{H}} \tilde{\mm{V}} \mm{P}^{1/2} \vv{d} + \mm{P}^{-1/2} \mm{\Delta} \vv{n}_{\mathrm{R}},
\end{equation}
where $\tilde{\mm{H}}=\left[ \mm{H}_{\mathrm{R}} \ \ -\mm{H}_{\mathrm{I}} \right]$ and $\tilde{\mm{V}} = \left[ \mm{V}_{\mathrm{R}}^\transp \ \ \mm{V}_{\mathrm{I}}^\transp \right]^\transp $. Here, $\mm{H}_{\mathrm{R}} / \mm{H}_{\mathrm{I}}$ and $\mm{V}_{\mathrm{R}} / \mm{V}_{\mathrm{I}}$ are the real/imaginary parts of $\mm{H}$ and $\mm{V}$, respectively. Furthermore, $\vv{n}_{\mathrm{R}}=\left[ n_{\mathrm{R}_1} \ldots n_{\mathrm{R}_K} \right]^\transp$ is the real part of the noise vector $\vv{n}$ with variance $\sigma_{n_\mathrm{R}}^2=0.5\sigma_{n}^2$. In Fig. \ref{fig:DL_BD}, the block diagram of the downlink augmented real-valued system model is shown. The design goal in this work is the optimization of $\tilde{\mm{V}}$ for the minimization of the sum MSE. The corresponding optimization problem can be formulated as
\vspace*{-2.2mm}
\begin{align}
\label{Eqn_DL_Opt}
& \min_{\tilde{\mm{V}} }  {\expect{{\Vert \vv{d} - \hat{\vv{d}} \Vert}^2}} \\
\text{subject~to:} \ \mathrm{tr} \left(\mm{P} \tilde{\mm{V}}^\herm \tilde{\mm{V}} \right) &= P_{\mathrm{TX}}, \ \ P_k \geq 0, \forall k \in \lbrace 1,\ldots,K \rbrace, \nonumber
\end{align}
where $P_{\mathrm{TX}}$ denotes the joint transmit power budget of all UTs. Because of the coupling of the different UTs introduced by the precoding matrix, the constrained downlink optimization problem in (\ref{Eqn_DL_Opt}) is difficult to solve. In contrast, in the uplink, each vector of the detection matrix can be optimized separately and the corresponding optimization problem is easier to solve. In the next section, we exploit results from uplink/downlink duality to transform the original downlink system into its equivalent uplink counterpart and solve the much simpler optimization problem in the uplink \cite{Shi2007}.
\vspace*{-0.4mm}

\begin{figure}[tp]
\begin{center}
\psfrag{p1}[cc][cc][.7]{$Q_1^{1/2}$}
\psfrag{pK}[cc][cc][.7]{$Q_K^{1/2}$}
\psfrag{d1}[cc][cc][1]{$d_1$}
\psfrag{dK}[cc][cc][1]{$d_K$}
\psfrag{x}[cc][cc][1]{$\vv{x}$}
\psfrag{U}[cc][cc][1]{$\tilde{\mm{U}}$}
\psfrag{sQ}[cc][cc][1]{$\ {\mm{Q}}^{1/2}$}
\psfrag{isP}[cc][cc][0.75]{$\ {\mm{Q}}^{-1/2}$}
\psfrag{s}[cc][cc][1]{$\vv{s}$}
\psfrag{GH}[cc][cc][1]{$\tilde{\mm{H}}^{\transp}$}
\psfrag{n}[cc][cc][1]{$\vv{n}_\mathrm{R}$}
\psfrag{r}[cc][cc][1]{$\vv{r}$}
\psfrag{Del}[cc][cc][1]{$\mm{\Delta}$}
\psfrag{dlK}[cc][cc][1]{$\delta_K$}
\psfrag{q1}[cc][cc][1]{$\sqrt{q_1}$}
\psfrag{qK}[cc][cc][1]{$\sqrt{q_2}$}
\psfrag{dd}[cc][cc][1]{$\hat{\vv{d}}$}
\includegraphics[width=\linewidth]{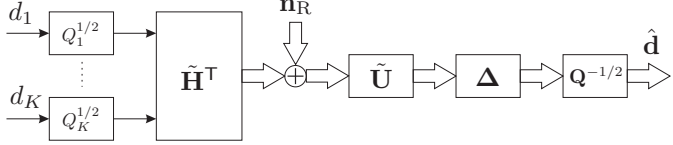}
\caption{Dual uplink augmented real-valued system model.}
\label{fig:UL_BD}
\end{center}
\vspace*{-6mm}
\end{figure}

%
\vspace*{-2mm}
\section{Widely-Linear Precoding}\label{Sec_WL_Precoder}
In this section, we use the uplink/downlink duality to derive the WL precoders. First, we derive the optimal WL-MMSE precoder in Section \ref{Sec_Opt_WL_Precoder}. Then, in Section \ref{Sec_PE_WL_Precoder}, we introduce the low-complexity PE WL-MMSE precoder.
\vspace*{-0.2mm}
\subsection{Optimal WL-MMSE Precoding}\label{Sec_Opt_WL_Precoder}
One of the main results of the uplink/downlink duality of sum MSE minimization in MU-MISO systems states that under the same sum power constraint, in the downlink system, the same sum MSE can be achieved as in the equivalent dual uplink system, if the power allocation, precoding, and detection matrices are chosen appropriately \cite{Shi2007}. For complex-valued system models, the dual uplink system model is obtained by adopting the Hermitian transposes of the downlink channel and precoding matrices as the uplink channel and detection matrices, respectively. Here, we extend this concept to augmented real-valued system models, cf. Fig. \ref{fig:UL_BD}. In the dual uplink system model, depicted in Fig. \ref{fig:UL_BD}, the sum MSE minimization problem can be formulated as
\vspace*{-2.5mm}
\begin{align}
\label{Eqn_UL_Opt}
& \min_{\tilde{\mm{U}} }  \expect{{\Vert \vv{d} - \hat{\vv{d}} \Vert}^2} \\
\hspace*{-11mm}\text{subject~to:} \ \mathrm{tr} \left(\mm{Q} \right) = & P_{\mathrm{TX}}, \ \ Q_k \geq 0, \forall k \in \lbrace 1,\ldots,K \rbrace, \nonumber
\end{align}
where $\hat{\vv{d}}$ can be expressed as
\vspace*{-0.4mm}
\begin{equation}
\hat{\vv{d}}= {\mm{Q}}^{-1/2} \mm{\Delta} \tilde{\mm{U}} {\tilde{\mm{H}}}^\transp {\mm{Q}}^{1/2} \vv{d} + {\mm{Q}}^{-1/2} \mm{\Delta} \tilde{\mm{U}} \vv{n}_{\mathrm{R}}. 
\end{equation}
Here, ${\mm{Q}}$ is the power allocation matrix and $\tilde{\mm{U}}$ is the normalized version of the detection matrix $\check{\mm{U}}= \mm{\Delta} \tilde{\mm{U}}$ with unit norm rows, where diagonal matrix $\mm{\Delta}$ contains the norms of the rows of $\check{\mm{U}}$. We define the signal-to-noise ratio (SNR) as $ \mathrm{SNR} \triangleq P_\mathrm{TX} / \sigma_n^2 $. To focus on the precoder design, we assume that all UTs transmit with equal powers, i.e., $\mm{Q}= \left(P_\mathrm{TX}/K\right) \mm{I}_K $. This makes the problem in \eqref{Eqn_UL_Opt} convex with $\tilde{\mm{U}} = {\mm{\Delta}}^{-1} \check{\mm{U}}$ as the optimum solution, where $\check{\mm{U}}$ is given by
\vspace*{-5mm}
\begin{align}
\check{\mm{U}} =  \frac{1}{N} \ \tilde{\mm{H}} \left( \frac{1}{N} \tilde{\mm{H}}^\transp \tilde{\mm{H}} + \frac{\sigma_{n_\mathrm{R}}^2}{P_\mathrm{TX}} \frac{K}{N} \mm{I}_{2N} \right)^{-1}.
\label{Eqn_WLMMSE_Detector}
\end{align}
\vspace*{-0.4mm}
Now, exploiting the uplink/downlink duality, the normalized precoding matrix is obtained as \cite{Shi2007}
\vspace*{-1mm}
\begin{align}
\tilde{\mm{V}} = \tilde{\mm{U}}^\transp.
\end{align}
We further use the uplink/downlink duality to obtain the dual counterpart of the power allocation in the downlink as $\mm{P} = \mathrm{diag} \left( \vv{p} \right)$, where vector $\vv{p}$ is given by \cite[Eq. (10.44)]{Tse2011},
\vspace*{-2mm}
\begin{equation}
\vv{p} = \frac{\sigma_{n}^2}{2} \left( \mm{I}_K - \mm{B}^\transp \right)^{-1} \vv{b}.
\end{equation}
\vspace*{-1.3mm}
Here, matrix $\mm{B}$ is defined as $\left[ \mm{B} \right]_{m, n} = \Big| { \big[ \tilde{\mm{H}} \big] }_{n,:} { \big[ \tilde{\mm{V}} \big] }_{:,m} \Big|^2$, and the elements of vector $\vv{b}=\left[b_1 \ldots b_K\right]^\transp$ are given by \cite{Tse2011}
\begin{equation}
b_k = \frac{\mathrm{SINR}_k^{\mathrm{UL}}}{\left(1+\mathrm{SINR}_k^{\mathrm{UL}} \right) \Big| { \big[ \tilde{\mm{H}} \big] }_{k,:} { \big[ \tilde{\mm{V}} \big] }_{:,k} \Big|^2 }.
\end{equation}
\vspace*{-0.4mm}
$\mathrm{SINR}_k^{\mathrm{UL}}$ is the SINR at the $k$th UT in the uplink and is defined as
\vspace*{-4mm}
\begin{equation}
\mathrm{SINR}_k^{\mathrm{UL}} \hspace*{-1.2mm} \triangleq \hspace*{-1.2mm} \frac{Q_k \Big| { \big[ \tilde{\mm{U}} \big] }_{k,:} { \big[ \tilde{\mm{H}}^\transp \big] }_{:,k} \Big|^2  }{0.5 \ \sigma_n^2 \big\| {\big[\tilde{\mm{U}} \big]}_{k,:} \big\| ^2 \hspace*{-1.5mm} + \hspace*{-1.0mm} \sum_{j=1 \atop j \neq k}^{K} Q_j \Big| { \big[ \tilde{\mm{U}} \big] }_{k,:} { \big[ \tilde{\mm{H}}^\transp \big] }_{:,j} \Big|^2}.
\end{equation}
\vspace*{-0.4mm}
Finally, the normalized WL precoding matrix $\mm{V}= \mm{V}_{\mathrm{R}} + j \mm{V}_{\mathrm{I}}$ can be constructed from its augmented version $\tilde{\mm{V}} = \left[ \mm{V}_{\mathrm{R}}^\transp \ \ \mm{V}_{\mathrm{I}}^\transp \right]^\transp $.
\vspace*{-3mm}
\subsection{PE WL-MMSE Precoding}\label{Sec_PE_WL_Precoder}
In the following, we derive the PE WL-MMSE precoder. To this end, we approximate the matrix inversion in the detector matrix $\check{\mm{U}}$ by a matrix polynomial and rewrite (\ref{Eqn_WLMMSE_Detector}) as
\vspace*{-2mm}
\begin{align}
\check{\mm{U}}_\mathrm{PE} = \frac{1}{N} \ \tilde{\mm{H}} \sum_{l=0}^{L} \omega_l \left( \frac{1}{N} \tilde{\mm{H}}^\transp \tilde{\mm{H}} \right)^l.
\label{Eqn_PEWLMMSE_Precoder}
\end{align}
\vspace*{-0.4mm}
Adopting the minimization of the average energy of the difference between the WL-MMSE detector’s output and the PE WL-MMSE detector’s output as the optimization objective, the optimal coefficients $\boldsymbol{\omega} = \left[ \omega_0 \ldots \omega_L \right]^\transp$ are calculated as $\boldsymbol{\omega} = \mm{\Xi}^{-1} \cdot \boldsymbol{\varphi}$, where the elements of matrix $\mm{\Xi}$ are given by \cite{Moshavi1999}
\vspace*{-3mm}
\begin{align}
\left[ \vv{\Xi} \right]_{m, n} = \xi^{\left(m+n\right)} + \frac{\sigma_{n_\mathrm{R}}^2}{P_\mathrm{TX}} \frac{K}{N} \xi^{\left(m+n-1\right)},
\label{Eqn16}
\end{align}
and the elements of vector $\boldsymbol{\varphi}$ are defined as $\left[ \boldsymbol{\varphi} \right]_m = \xi^{\left(m\right)}$.
Here, $\xi^{\left(m\right)}$ is the $m$th order moment of the eigenvalues of the large matrix $\frac{1}{N} \tilde{\mm{H}}^\transp \tilde{\mm{H}}$ and given by \cite[Theorem 1]{Mueller2001},
\vspace*{-3mm}
\begin{align}
\xi^{\left(m\right)} \xrightarrow[K, N\rightarrow\infty]{\mathrm{a.s.}} \sum_{i=0}^{m-1} \binom{m}{i} \binom{m}{i+1} \frac{ (\beta/2)^i}{m}.
\label{Eqn18}
\end{align}
\vspace*{-0.4mm}
The optimal coefficient vector $\boldsymbol{\omega}$ can be calculated easily and does not depend on the instantaneous realizations of $\tilde{\mm{H}}$. The PE WL-MMSE precoder is then obtained by replacing $\check{\mm{U}}$ with $\check{\mm{U}}_\mathrm{PE}$ in Section \ref{Sec_Opt_WL_Precoder}. Exploiting the structure of (\ref{Eqn_PEWLMMSE_Precoder}) and applying Horner's scheme, the PE WL-MMSE precoded data vectors can be calculated by performing matrix-vector multiplications only while avoiding matrix-matrix multiplications, see \cite{Zarei_PIMRC2013}, \cite{Mueller2001} for details. This leads to a computational complexity of $ \mathcal{O} \left( KN \right)$ for calculation of one precoded data vector.

%

\vspace*{-3mm}
\section{Large System Analysis}\label{Sec_LargeSystem}
In this section, we use results from random matrix theory to derive asymptotic expressions for the UTs' SINRs and the sum rate in the downlink of a large-scale MU-MISO system. Since the SINRs in the downlink and the dual uplink system are identical \cite{Shi2007} and due to the fact that large system analysis of the detector in the uplink is simpler than analysis of the downlink precoder, we derive the asymptotic SINRs for the dual uplink model. 
First, we analyze the SINRs in the uplink for conventional MMSE detection. The corresponding detected signal of the $k$th UT can be expressed as
\vspace*{-2mm}
\begin{align}
{\hat{d}}^{\mathrm{MMSE}}_k &= \left[ \mm{H} \right]_{k,:} \big( \mm{H}^\herm \mm{H} + \frac{\sigma_n^2}{\rho} \mm{I}_N \big)^{-1} \left[ \mm{H} \right]^\herm_{k,:} d_k \nonumber \\
& + \sum_{j \neq k}^{K} \left[ \mm{H} \right]_{k,:} \big( \mm{H}^\herm \mm{H} + \frac{\sigma_n^2}{\rho} \mm{I}_N \big)^{-1} \left[ \mm{H} \right]^\herm_{j,:} d_j \nonumber \\
& + \frac{1}{\sqrt{\rho}} \left[ \mm{H} \right]_{k,:} \big( \mm{H}^\herm \mm{H} + \frac{\sigma_n^2}{\rho} \mm{I}_N \big)^{-1} \vv{n},
\label{UL_SysMod}
\end{align}
where $\rho=P_\mathrm{TX}/K$. Now, we define the following variables
\vspace*{-0.4mm}
\begin{align}
\xi_k \triangleq \lim_{K, N \rightarrow \infty} \frac{1}{N} \left[ \mm{H} \right]_{k,:} \left( \frac{1}{N} \mm{H}_k^\herm \mm{H}_k + \gamma \mm{I}_N \right)^{-1} \left[ \mm{H} \right]^\herm_{k,:}
\label{Formula_xi}
\end{align}
\vspace*{-0.4mm}
\begin{align}
\psi_k \triangleq \lim_{K, N \rightarrow \infty} \frac{1}{N} \left[ \mm{H} \right]_{k,:} \left( \frac{1}{N} \mm{H}_k^\herm \mm{H}_k + \gamma \mm{I}_N \right)^{-2} \left[ \mm{H} \right]^\herm_{k,:}
\label{Formula_psi}
\end{align}
\vspace*{-0.4mm}
\begin{align}
\zeta_k \triangleq & \lim_{K, N \rightarrow \infty} \frac{1}{N} \left[ \mm{H} \right]_{k,:} \left( \frac{1}{N} \mm{H}_k^\herm \mm{H}_k + \gamma \mm{I}_N \right)^{-1} \frac{1}{N} \mm{H}_k^\herm \mm{H}_k \nonumber \\
& \times \big( \frac{1}{N} \mm{H}_k^\herm \mm{H}_k + \gamma \mm{I}_N \big)^{-1} \left[ \mm{H} \right]^\herm_{k,:},
\label{Formula_zeta}
\end{align}
where $\gamma = \sigma_n^2 / (\rho N) $ and $\mm{H}_k$ is identical to matrix $\mm{H}$ with the $k$th row removed. Here, $\xi_k$, $\gamma \psi_k$, and $\zeta_k$ are the asymptotic values of the magnitude of the useful signal, noise power, and interference power of the $k$th UT for $K, N \rightarrow \infty$, respectively. Using the above defined variables, the asymptotic value of the SINR of the $k$th UT for $K, N \rightarrow \infty$ can be expressed as
\vspace*{-3mm}
\begin{align}
\mathrm{SINR}^{\mathrm{MMSE}^\circ}_k \left( \beta, \gamma \right) = \frac{\xi_k^2}{\zeta_k + \gamma \psi_k}.
\label{EqnSINR}
\end{align}
\vspace*{-0.4mm}
Now, exploiting \cite[Corollary 1]{Evans2000} yields
\vspace*{-0.4mm}
\begin{align}
&\xi_k = \lim_{K, N \rightarrow \infty} \frac{1}{N} \left[ \mm{H} \right]_{k,:} \Big( \frac{1}{N} \mm{H}_k^\herm \mm{H}_k + \gamma \mm{I}_N \Big)^{-1} \left[ \mm{H} \right]^\herm_{k,:} \nonumber \\
& = \mathrm{tr} \left( \frac{1}{N} \mm{H}_k^\herm \mm{H}_k + \gamma \mm{I}_N \right)^{-1}= \int_{-\infty}^{\infty} \frac{dF_{\mm{\Lambda}}\left( s \right)}{s+\gamma} \triangleq H_{\mm{\Lambda}}\left(\beta, -\gamma \right) \nonumber \\
& = \sqrt{\frac{\left(1-\beta\right)^2}  {4\gamma^2} + \frac{\left(1+\beta\right)}{2\gamma}+\frac{1}{4}} + \frac{1-\beta}{2\gamma} - \frac{1}{2}, 
\label{EqnStieltjes}
\end{align} 
where $dF_{\mm{\Lambda}}\left( s \right)$ is the empirical distribution of the eigenvalues of $\frac{1}{N} \mm{H}_k^\herm \mm{H}_k = \mm{T} \mm{\Lambda} \mm{T}^\herm$ with $\mm{\Lambda}$ and $\mm{T}$ being the matrix of eigenvalues and the matrix of eigenvectors, respectively. Here, the Stieltjes transform of $dF_{\mm{\Lambda}}\left( s \right)$ is denoted by $H_{\mm{\Lambda}} \left(\beta, \lambda \right)= \int_{-\infty}^{\infty} {(s-\lambda)}^{-1} dF_{\mm{\Lambda}} (s)$. Using \cite[Corollary 1]{Evans2000}, applying the above mentioned eigen-decomposition, and considering $\mm{T} \mm{T}^\herm= \mm{I}_N$, the following expression is obtained for $\psi_k$
\vspace*{-1mm}
\begin{align}
& \psi_k = \lim_{K, N \rightarrow \infty} \frac{1}{N} \left[ \mm{H} \right]_{k,:} \Big( \frac{1}{N} \mm{H}_k^\herm \mm{H}_k + \gamma \mm{I}_N \Big)^{-2} \left[ \mm{H} \right]^\herm_{k,:} =  \nonumber \\
& \mathrm{tr} \bigg( \Big( \frac{1}{N} \mm{H}_k^\herm \mm{H}_k + \gamma \mm{I}_N \Big)^{-2} \bigg) \xrightarrow[K, N\rightarrow\infty]{\mathrm{a.s.}} \mathrm{tr} \bigg( \Big( \mm{\Lambda} + \gamma \mm{I}_N \Big)^{-2} \bigg) \nonumber \\
& \xrightarrow[K, N\rightarrow\infty]{\mathrm{a.s.}} \int_{-\infty}^{\infty} \frac{dF_{\mm{\Lambda}} \left( s \right)}{\left(s+\gamma\right)^2}=-\frac{\partial H_{\mm{\Lambda}}\left(\beta, -\gamma\right)}{\partial \gamma}.
\label{EqnPsi2}
\end{align}
\vspace*{-0.4mm}
Using a similar procedure and performing algebraic operations, $\zeta_k$ can be expressed as \cite{Evans2000}, \cite{Nguyen2008}
\vspace*{-0.4mm}
\begin{align}
& \zeta_k = \lim_{K, N \rightarrow \infty} \frac{1}{N} \left[ \mm{H} \right]_{k,:} \bigg( \frac{1}{N} \mm{H}_k^\herm \mm{H}_k + \gamma \mm{I}_N \bigg)^{-1} \frac{1}{N} \mm{H}_k^\herm \mm{H}_k \times \nonumber \\
& \bigg( \frac{1}{N} \mm{H}_k^\herm \mm{H}_k + \gamma \mm{I}_N\bigg)^{-1} \left[ \mm{H} \right]^\herm_{k,:} \xrightarrow[K, N\rightarrow\infty]{\mathrm{a.s.}} \mathrm{tr} \left( \mm{\Lambda} \left( \mm{\Lambda} + \gamma \mm{I}_N \right)^{-2} \right) \nonumber \\
& = \int_{-\infty}^{\infty} \frac{sdF_{\mm{\Lambda}}\left( s \right)}{\left(s+\gamma\right)^2} = \int_{-\infty}^{\infty} \frac{dF_{\mm{\Lambda}} \left( s \right)}{s+\gamma} -\gamma \int_{-\infty}^{\infty} \frac{dF_{\mm{\Lambda}}\left( s \right)}{\left(s+\gamma\right)^2} \nonumber \\
& = H_{\mm{\Lambda}}\left(\beta, -\gamma\right) + \gamma \frac{\partial}{\partial \gamma} H_{\mm{\Lambda}}\left(\beta, -\gamma\right).
\label{EqnZeta2}
\end{align}
\vspace*{-0.5mm}
Substituting (\ref{EqnStieltjes})-(\ref{EqnZeta2}) into (\ref{EqnSINR}) yields the asymptotic SINR of the $k$th UT in the uplink. Since this SINR is identical to the SINR of the $k$th UT in the dual downlink system, the asymptotic sum rate in the downlink with MMSE precoding and complex-valued Gaussian data symbols is given by
\vspace*{-3mm}
\begin{align}
R_{\mathrm{MMSE}}^\circ = \sum_{k=1}^{K} \log_2 \left( 1 + \mathrm{SINR}^{\mathrm{MMSE}^\circ}_k \left( \beta, \gamma \right) \right).
\end{align}
Now, we are ready to provide the uplink SINR for WL-MMSE detection.\\
\emph{Theorem 1}: The asymptotic SINR of the $k$th UT in the uplink of a MU-MISO system with $K, N \rightarrow \infty$ using real-valued transmit symbols and WL-MMSE detection is given by  $\mathrm{SINR}^{\mathrm{WL-MMSE}^\circ}_k \left( \beta, \gamma \right) = \mathrm{SINR}^{\mathrm{MMSE}^\circ}_k \left( \beta/2, \gamma/2 \right)$.\\
\begin{IEEEproof}
See Appendix.
\vspace*{-0.4mm}
\end{IEEEproof}
Using the above theorem and the uplink/downlink duality, the sum rate of the downlink system using real-valued Gaussian data symbols and a WL-MMSE precoder can be expressed as
\vspace*{-4mm}
\begin{align}
R_{\mathrm{WL-MMSE}}^\circ = 0.5 \sum_{k=1}^{K} \log_2 \left( 1 + \mathrm{SINR}^{\mathrm{MMSE}^\circ}_k \left( \beta/2, \gamma/2 \right) \right). \nonumber
\end{align}
%
\vspace*{-6mm}
\section{Numerical Results}\label{Sec_Results}
\begin{figure}[tbp]
\begin{center}
\psfrag{rho}[cc][cc][0.6]{$\mm{Q}=\rho \mm{A}^{-1}$}
\includegraphics[width=\linewidth, clip=true]{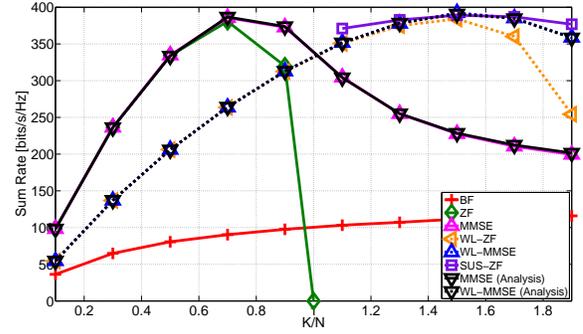}
\caption{\linespread{2} \small Sum rate vs. $K/N$ for $\mathrm{SNR}=$ 20 dB, $N=100$.}
\label{Rate_vs_Beta}
\end{center}
\vspace*{-6mm}
\end{figure}

\vspace*{-1mm}
In order to evaluate the performance of the proposed WL precoder, Monte-Carlo simulations have been conducted. 
The noise variance is assumed to be $\sigma_n^2=1$. In Fig. \ref{Rate_vs_Beta}, the ergodic sum rates of the MMSE, ZF, conjugate beamforming (BF), WL-MMSE, SUS-ZF \cite{Yoo2006}, and WL-ZF precoders for SNR = 20 dB and $N=100$ base station antennas are depicted. Conjugate BF is the downlink counterpart of the matched filter in the uplink. The ergodic sum rate is given by $R=\sum_{k=1}^{K} \mathbb{E} \lbrace \mathrm{log}_2 \left( 1+\mathrm{SINR}_k \right) \rbrace$, where the expectation is approximated by averaging over a sufficient number of channel realizations. The WL-ZF precoding matrix is obtained by setting the detector matrix in the dual uplink model to $\check{\mm{U}} = ( \tilde{\mm{H}} \tilde{\mm{H}}^\transp )^{-1} \tilde{\mm{H}}$ and using the procedure described in Section \ref{Sec_WL_Precoder} to obtain the precoder. 

As can be seen in Fig. \ref{Rate_vs_Beta}, with increasing $K/N$, the difference in performance between conjugate BF and the other schemes increases until the load factor reaches $K/N=0.7$, where the MMSE precoder achieves the highest sum rate performance among the considered schemes. For $K/N < 1$, the MMSE precoder outperforms the WL-MMSE precoder. This is due to the fact that for $K<N$, the base station has enough spatial degrees of freedom to efficiently suppress interference from $K-1$ users if complex transmit symbols and MMSE precoding are employed. On the other hand, for $K<N$, the sum rate of the WL-MMSE precoder is compromised by the waste of dimensions caused by the limitation to real-valued transmit symbols.
For $K>N$, the WL-MMSE precoder achieves a significantly higher sum rate compared to the conventional MMSE precoder. This occurs because the WL-MMSE precoder employs real-valued transmit symbols, which enables it to relegate the interference to the imaginary part of the received signal, making it invisible to the receiver that inspects only the real part of the observation.
In addition, in contrast to the WL-ZF precoder's sum rate, which decreases significantly for $K/N>1.5$, the sum rate of the WL-MMSE precoder is almost constant for $1.5<K/N<1.9$. In fact, the proposed WL-MMSE precoder closely approaches the sum rate of the SUS-ZF precoder \cite{Yoo2006}. 
Moreover, in contrast to the SUS-ZF precoder, where UTs with poor channels are allocated zero rate, with the proposed WL-MMSE precoder, always all UTs are served. Furthermore, in Fig. \ref{Rate_vs_Beta}, we also present analytical results for the sum rate obtained from the large system analysis for the conventional MMSE and WL-MMSE precoders. A perfect match between analytical results and simulation results is observed. 

\begin{figure}[tbp]
\begin{center}
\psfrag{rho}[cc][cc][0.6]{$\mm{Q}=\rho \mm{A}^{-1}$}
\includegraphics[width=\linewidth, clip=true]{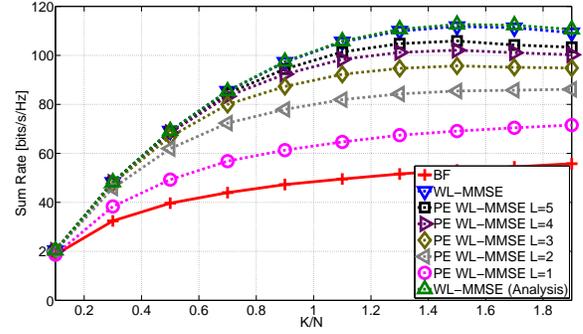}
\caption{\linespread{2} \small Sum rate vs. $K/N$ for $\mathrm{SNR}=$ 15 dB, $N=50$.}
\label{Rate_vs_Beta_PE}
\end{center}
\vspace*{-6mm}
\end{figure}

In Fig. \ref{Rate_vs_Beta_PE}, the sum rates of PE WL-MMSE precoders with different polynomial orders $L$ are compared to the sum rate of the BF and WL-MMSE precoders for SNR = 15 dB and $N=50$. As can be observed, for increasing $L$, the PE WL-MMSE precoder approaches the sum rate of the WL-MMSE precoder. For example, for $L=4$ and $K/N=1.5$, the PE WL-MMSE precoder achieves almost $91\%$ of the WL-MMSE precoder's sum rate and thereby also approaches the sum rate of the SUS-ZF precoder. However, for $K>N$, the computational complexity of calculating one precoding vector for PE WL-MMSE precoding and SUS-ZF precoding is $\mathcal{O} \left( KN \right)$ and $\mathcal{O} \left( KN^2 \right)$ \cite{Yoo2006}, respectively, i.e., PE WL-MMSE entails a lower complexity.
\vspace*{-0.4mm}
%

%
\vspace*{-3mm}
\appendices
\section*{Appendix - Proof of Theorem 1}
The detected signal in the uplink of a MU-MISO system using WL-MMSE detection is given by (\ref{UL_SysMod}), but with $\mm{H}$, $\vv{n}$, and $\sigma_n^2$ being replaced by $\tilde{\mm{H}}$, $\vv{n}_{\mathrm{R}}$, and $0.5 \ \sigma_n^2$, respectively. Thus, for the WL-MMSE detector, a similar SINR expression as for the conventional MMSE detector results. Furthermore, both matrices $\tilde{\mm{H}}$ and ${\mm{H}}$ have zero-mean i.i.d. Gaussian distributed entries, but the dimension of $\tilde{\mm{H}}$ is $K \times 2N$ whereas that of $\mm{H}$ is $K \times N$. Therefore, the Stieltjes transform of $dF_{\frac{1}{N}\tilde{\mm{H}}^\herm \tilde{\mm{H}}} \left( s \right)$ is obtained by replacing $\beta$ in the Stieltjes transform of $dF_{\frac{1}{N}\mm{H}^\herm \mm{H}} \left( s \right)$ by $K / \left(2N\right)= \beta/2$. Moreover, the SINRs in the uplink system using WL-MMSE and MMSE detection are only functions of the Stieltjes transform of $dF_{\frac{1}{N}\tilde{\mm{H}}^\herm \tilde{\mm{H}}} \left( s \right)$ and $dF_{\frac{1}{N}\mm{H}^\herm \mm{H}} \left( s \right)$, and their derivative with respect to $\gamma$, respectively. In addition, we have $\sigma_{n_\mathrm{R}}^2=0.5\sigma_n^2$. Hence, the SINR in the uplink system using WL-MMSE detection is obtained by replacing $\beta$ with $\beta/2$ and $\gamma$ with $\gamma/2$ in the SINR expression of the uplink system using MMSE detection. 
\vspace*{-4mm}

\ifCLASSOPTIONcaptionsoff
  \newpage
\fi

\bibliographystyle{IEEEtran}
\bibliography{Massive_MIMO}

\end{document}